# Revealing Correlation of Valence State with Nanoporous Structure in Cobalt Catalyst Nanoparticles by *in Situ* Environmental TEM

*Huolin L. Xin[1], Elzbieta A. Pach[1], Rosa E. Diaz[2], Eric A. Stach[2], Miquel Salmeron[1], Haimei Zheng[1]\**

[1]Materials Sciences Division, Lawrence Berkeley National Laboratory, Berkeley, CA 94720, United States

[2]Center for Functional Nanomaterials, Brookhaven National Laboratory, Upton, NY 11973, United States

*Corresponding author: hmzheng@lbl.gov

ABSTRACT

Simultaneously probing the electronic structure and morphology of materials at the nanometer or atomic scale while a chemical reaction proceeds is significant for understanding the underlying reaction mechanisms and optimizing a materials design. This is especially important in the study of nanoparticle catalysts, yet such experiments have rarely been achieved. Utilizing an environmental transmission electron microscope (ETEM) equipped with a differentially pumped gas cell, we are able to conduct nanoscopic imaging and electron energy loss spectroscopy (EELS) *in situ* for cobalt catalysts under reaction conditions. Analysis revealed quantitative correlation of the cobalt valence states to the particles' nanoporous structures. The *in situ* experiments were performed on nanoporous cobalt particles coated with silica while a 15 mTorr hydrogen environment was maintained at various temperatures (300-600 ºC). When the nanoporous particles were reduced, the valence state changed from cobalt oxide to metallic cobalt and concurrent structural coarsening was observed. I*n situ* mapping of the valence state and the corresponding nanoporous structures allows quantitatively analysis necessary for understanding and improving the mass activity and lifetime of cobalt-based catalysts, *i.e.*, for Fischer-Tropsch synthesis that converts carbon monoxide and hydrogen into fuels, and uncovering the catalyst optimization mechanisms.

KEYWORDS: Environmental TEM, *in situ* TEM, Cobalt catalysts, porosity control, Fischer–Tropsch synthesis

Accompanying recent accomplishments in nanocharacterization, the electronic structure or morphology of materials can be studied *in situ* at the nanometer or atomic scale while chemical reactions are proceeding. For example, ambient pressure X-ray absorption spectroscopy provides electronic structures of ensembles of catalyst nanoparticles during catalytic reactions.[1, 2] Environmental transmission electron microscopy (TEM) has been developed to image single nanoparticle growth dynamics in liquids[3] or catalytic nanoparticle-gas reactions in real time under relevant catalytic conditions.[4] However, there have been limited studies on simultaneous *in situ* imaging and spectroscopy at the nanoscale, which is key in revealing the correlation between the valence state and morphology changes of active-metal-containing nanoparticles, *i.e.*, during heterogeneous catalysis. With the development of gas environmental TEM[5] and electron energy loss spectroscopy (EELS), the valence state of nanoparticles can be measured concurrently with atomic-resolution imaging during the reaction. By taking advantage of this development, we study the valence state of nanoporous cobalt-containing particles and its correlation with the structural coarsening during the hydrogen reduction reaction for Fischer-Tropsch (F-T) synthesis—an industrial reaction that converts syngas (a mixture of hydrogen and carbon monoxide) to liquid fuels.[6, 7]

The initial reduction of oxide nanoparticles into metallic catalysts is a critical step in Fischer-Tropsch (F-T) synthesis. Among the various active metal catalysts (Ni, Co, Fe and Ru), iron and cobalt are the only catalysts that are used in commercial F-T reactors, as they exhibit both low cost and high selectivity, with cobalt being preferred for the synthesis of heavy hydrocarbons such as jet and diesel fuels.[8-10] The preparation and conditioning of the microstructure and valence state of the catalyst are essential for achieving the required durability and catalytic activity. Because cobalt oxides are typically used for catalyst formation, a crucial step of catalyst preparation is the hydrogen reduction of oxides at high temperatures. However, during this process sintering of the catalyst particles reduces the porosity and thus the accessible surface area. This problem can be partially mitigated by mixing the catalysts with silica to form self-assembled composites.[11] Previous studies have shown that partially reduced F-T catalysts can be as active as a fully reduced one.[8, 12] Therefore, an understanding of the correlation between the valence state and the underlying nanoporous structure of the catalysts is critical to yield an optimized catalyst. In this work, we use the state-of-the-art aberration-corrected environmental transmission electron microscopy (ETEM) and electron energy loss spectroscopy (EELS) to make this connection between nanoporous structure and the valence state of cobalt/silica catalysts during hydrogen reduction. Such *in situ* correlation provides insights for future optimization of F-T catalysts in particular, and more broadly yields insights into porosity control in the general class of metal/metal-oxide nanocomposites.

The $CoO_x$/Silica nanocomposites were synthesized using a two-step method.[13] First, Co nanoparticles (10-12 nm) were prepared using air-free colloidal synthesis. Subsequently, a silica shell was formed around the self-assembled Co clusters by injection of tetramethyorthosilicate (TMOS) and octadecyltrimethoxysilane ($C_{18}$TMS) into the nanoparticle suspension. The as-prepared $CoO_x/SiO_2$ was dispersed in isopropanol and spread on a nonporous amorphous silicon TEM membrane (SIMpore, 5 nm amorphous Si with 100 μm windows) by micro-pipetting. The surfaces of the membranes were passivated by native oxidation in air, as determined by EELS (data not shown). *In situ* reduction, imaging and chemical analysis were performed using an FEI environmental-cell Titan 80/300 equipped with a post-specimen aberration corrector (CEOS) operated at 300 kV. This instrument utilizes a differential pumping environmental cell,[5] which allows injection of up to several Torr of gas during imaging, depending on the molecular weight of the gas. In the case of hydrogen, a maximum pressure of 1 Torr can be achieved. The gas pressure is controlled manually by a simple needle valve and is monitored by a high-accuracy pressure transducer. The ultrahigh purity [99.9999%] hydrogen used in this experiment was supplied through stainless steel tubes. Prior to the in-situ experiment, aberration coefficients were measured using a Zemlin tableau, and corrected until a quarter wavelength semi-angle larger than 20 mrad was achieved. Electron energy loss spectra were recorded with a Tridiem Gatan imaging filter with an energy dispersion of 0.3 eV/channel. The EELS spectra were acquired in selected-area TEM imaging mode (10 μm selected area aperture) with an energy resolution of ~2 eV. Following each core-loss spectrum acquisition, the spectrum of the elastically scattered electrons (zero-loss peak) was also recorded for energy axis calibration. A Gatan 652 Inconel double-tilt furnace-type heating holder was used for *in situ* heating. *Ex situ* studies of the as-synthesized samples and of the samples following *in situ* reduction were performed using a 200 kV FEI monochromated F20 UT Tecnai equipped with a Gatan imaging filter.

## RESULTS AND DISCUSSION

Figure 1a presents bright-field TEM images of the as-prepared $CoO_x/SiO_2$ nanocomposites dispersed on a carbon grid. Based on the image contrast and spatially-resolved EELS spectra (Figure 1c, 1d and S1), we can identify the cobalt oxide core and the silica shell. We further found that the majority of the core material is cobalt monoxide with a trace amount of $Co_3O_4$ (Fig. S2-4). The spatial variation in the bright-field image contrast within each nanocomposite particle (Fig. 1a and 1b) suggests that the cobalt-oxide core is likely porous, resulting from the sintering of multiple CoO crystalline nanoparticles. As shown in Figure S2, individual particles in the sintered core are separated by low-angle grain boundaries. However, projection images can sometimes be misleading. For instance, it is difficult to distinguish a porous network from a corrugated solid structure in a single projection image. In addition, the observed low-spatial-frequency contrast modulation can also be a result of diffraction contrast caused by strain fields. To reliably visualize the 3-D internal structure without ambiguity, we used annular dark-field electron tomography.[14-16] We recorded 73 annular dark-field scanning transmission electron microscopy (ADF-STEM) images of a single $CoO_x/SiO_2$ nanoporous particle from -72 degrees to 72 degrees with two degree intervals (see supplemental materials for more details). The 3-D structure of the material was reconstructed using the simultaneous iterative reconstruction technique (SIRT).[14] Figure 2 shows the 3-D rendering of the high-Z $CoO_x$ component in the nanocomposite using isosurfaces and progressive cross sections. These results clearly demonstrate in 3-D that the cobalt-oxide core is an interconnected nanoporous network. It is expected that this porous structure can facilitate the infiltration of gas molecules at reaction conditions. Figure 3a shows the heating and the hydrogen injection trajectories used in *in situ* experiments (see more details in Fig. S6). First, the sample was heated to 300 °C in vacuum and was maintained at the temperature for an hour to minimize thermal drift of the sample. No significant changes of the nanoporous core structure were observed during this stage (Fig. 3b(I) and Fig. S7(I)). Subsequently, 15 mTorr $H_2$ was injected into the environmental cell, leading to noticeable changes of the nanoporous core (Fig. 3b(II) and Fig. S7(II)). The structural changes became more evident when the temperature was raised to 450 °C (0.25 hour). As shown in Fig. 2b(III) and Fig. S7(III), the nanoporous structures coarsened significantly, although some degree of porosity was still preserved. Upon increasing the temperature to 500 °C (1.5 hours), morphological changes lead to particles which have limited porosity (Fig. 3b(IV) and Fig. S7(IV)). At 600 °C (0.5 hour), even larger cobalt particles with smooth surfaces in silica pockets were observed. During the sintering process, silica clearly provides certain degree of protection for the cobalt cores from inter-pocket sintering. However, in places with high-density nanocompsite aggregates, silica from different pockets can "glue" together to form a micron scale composite (Fig. S7(III, IV, V)). In individual isolated nanocomposites, silica shells predominantly follow the shape change of the core to form a core-shell structure.

To quantify the temperature-dependent structural evolution during hydrogen reduction, we measured the effective particle size of the porous network at different temperatures. Here, the effective particle size is the average internal feature size inside the core, which describes a porous network in terms of equivalent particles. Figure 4a shows the effective particle size as a function of reduction temperature. The increase of the effective particle size reflects coarsening of the network and a decrease of the surface-to-volume ratio at higher reduction temperatures. It demonstrates the evolution of the core structure from an initial nanoporous network towards solid spheres.

To correlate the structural coarsening with the fraction of reduced metallic cobalt within the core, we used *in situ* EELS to monitor the electronic structure changes of cobalt. Figure 4b shows the near edge fine structures of Co $L_{2,3}$ edges ($2p^6 3d^n$ -> $2p^5 3d^{n+1}$ transitions) recorded at four conditions (300 ℃+Vaccum, 300 ℃+$H_2$, 500 ℃+$H_2$ and 600 ℃+$H_2$ corresponding to I, II, IV, V in Figure 3a). Because the as-prepared material is mostly CoO, the 300 ℃+Vaccum spectrum can be approximately assigned as a $Co^{"2+"}$ fingerprint. Similarly, the 600 ℃+$H_2$ spectrum can be considered approximately fully reduced, *i.e.*, $Co^{"0"}$ fingerprint, since no obvious oxide diffraction reflections are present in the SAD pattern after reduction at 600 ℃ (Fig. S8). The $L_{2,3}$ near edge fine structures of cobalt with an average valence state in between can be decomposed into a linear combination of the two fingerprints.[17] The corresponding decomposition coefficient of the Co"0" component reflects the reduction fraction. Figure 4c shows such decompositions as a function of reduction temperature. We see that at 300 °C+$H_2$, 40±9% of the material is reduced. At 500 °C, only a small amount of residual cobalt oxide (5.5±3%) is present. In addition to demonstrating a general correlation between the increased reduction of cobalt and enhanced sintering, these results provide the first direct quantification between morphological changes and changes in electronic structures. We further calculated the optimum reduction conditions, based on the evolution of the surface-to-volume ratio and the reduction fraction, presented in Figure 4a-c. This shows that the optimum reduction temperature is within 390-440 °C in these experiments (Figure 4d), which is consistent with the optimal reduction temperature used in commercial F-T plants (~400 °C[8]). This study reveals the underlying mechanisms quantitatively for the first time by *in-situ* TEM.

It is noted that electron beams can cause knock-on damage, local heating and induced coalescence. For this reason we have used a nearly parallel beam with relative low intensity and avoided constant illumination of the materials by closing the gun valve. Most importantly, the particles that had never been exposed to the electron beam show a similar morphology to that of particles exposed to the beam (Figure S9). This demonstrates that there are minimal electron beam modifications in the *in situ* studies.

After full reduction of cobalt at 600 °C+$H_2$, we switched off the heating and stopped hydrogen flow after the sample had cooled to room temperature. The environmental cell was then pumped down to $10^{-6}$ torr and the reduced sample was left in the cell overnight. After 14 hours, we reinvestigated the reduced products. We found most of the particles were intact (Figure S10), but the particles with exposed surfaces had formed a layer of native oxide (Figure 5a and 5b). However, at room temperature, those particles with a 2-4 nanometer thick silica shell showed a high resistance to oxidation at ambient conditions. Figure 5 shows an *ex situ* images of such particles recorded three weeks after the *in situ* reduction experiment (also see Figure S10). The atomic resolution images suggest the surface is not oxidized. This was confirmed by the measurements of valence state at the catalyst surfaces using atomic-scale STEM-EELS. It is well known that the intensity ratio between the L2 and L3-edges is sensitive to the oxidation state of the metal[18-21]. However, the traditional L3/L2 method requires spectra with relatively good signal-to-noise ratio. To visualize the valence change of cobalt more reliably at curved surfaces, we directly integrate the L2 and L3 intensity on the background subtracted spectrum (Figure 5d). To validate our method, we first applied it to the particle with an oxidized surface, shown in Figure 5b and S12a. Figure 5e(II) shows the L3/L2 ratio as the electron probe scans from the native oxide layer to the metallic part of the particle. As expected, the oxide layer shows a significantly higher L3/L2 ratio than that of the metallic part. In Fig. 5f, we use this method to probe the particle shown in Fig. S12b. The L3/L2 line profile shows there is no statistical significant valence change from the surface to the center of the particle. This indicates the silica layer can readily allow gas molecules such as $H_2$ to penetrate through at elevated temperatures. However, a two-nanometer silica shell acts as an $O_2$ blocker at room temperature, protecting cobalt from oxidation at ambient conditions, even for extended periods of time.

## CONCLUSIONS

In conclusion, we performed *in situ* environmental TEM study of nanoporous cobalt/silica catalysts relevant to Fischer-Tropsch synthesis. We have demonstrated quantitatively that $H_2$ reduction of the $CoO_x$ into metallic cobalt results in an increase of an effective particle size of the porous structure. An optimum reduction temperature of about 410 ℃ was achieved at the current reduction environment. The correlation between the valence state and the structural changes provided here may be proven to be significant for the design of catalysts with improved catalytic activity and selectivity.

## MATERIALS AND METHODS

**Microscope.** All *in situ* imaging and spectroscopy were performed using a 300 kV image aberration-corrected environmental TEM at Brookhaven National Lab. *Ex situ* scanning/transmission electron microscopy (S/TEM) imaging and electron energy loss spectroscopy (EELS) were performed mainly using a 200 kV FEI Tecnai at National Center for Electron Microscopy of Lawrence Berkeley National Lab (LBNL). Some high-resolution TEM images were taken using a 300 kV image aberration-corrected environmental TEM at Brookhaven National Lab (Fig. 1b, Fig. S2, Fig. S7, Fig. S9e, Fig. S10a, Fig. S11a). Part of the bright-field TEM images and selected-area diffraction (SAD) patterns were recorded using a 200 kV JEOL 2100 at Materials Sciences Division of LBNL (Fig. 1a, Fig. S8, Fig. S11b).

- **Annular dark-field STEM imaging in Tecnai**
    - Schottky field emission gun
    - 200 kV
    - Semi-convergence angle: 11 mrad
    - Collection semi-angles: 26.4-100 mrad
- **Electron energy loss spectroscopy in Tecnai**
    - Semi-convergence angle: 11 mrad
    - Collection semi-angles: 0-22 mrad
    - Energy resolution: 0.5-1.5 eV depending on dispersion settings.
- **TEM imaging in JEOL 2100**
    - LaB6 cathode

o   200 kV

        o   Information transfer up to 1.8 angstrom

- **TEM imaging in BNL ETEM**

        o   Schottky field emission gun

        o   300 kV

        o   Quarter wavelength semi-angle better than 20 mrad

        o   Information transfer: sub-angstrom

**Annular Dark-Field Scanning Transmission Electron Tomography.** Annular dark-field STEM (ADF-STEM) images were recorded from -72 to 72 degrees at two-degree intervals. Beam current was around 10 pA (200 kV). Each image (1024x1024 pixels, field of view 84 nm) was acquired for 20 seconds. No mass loss was observed during the whole image acquisition process. The theoretical resolution for this reconstruction is ~2.2 nm. However, the resolution can be improved using the simultaneous iterative reconstruction algorithm.

The acquired tilt series was first aligned using the center of mass. The fine adjustment was made manually using a Matlab script package (e⁻Tomo) written by Robert Hovden at Cornell University. The 3-D dataset was reconstructed by the simultaneous iterative reconstruction algorithm implemented in Matlab. The script was initially written by one of the authors—HLX. It was modified and integrated into the e⁻Tomo package by Robert Hovden. 25 iterations were used for the final reconstructions.

Due to the limited tilt range (-72 to 72 degrees), a wedge of information is missing in reciprocal space of the reconstruction. This results in an elongation of the reconstructed features by a factor 1.27 along the beam incident direction.

***In situ* heating and gas injection.** *In situ* heating was achieved by a furnace-type double-tilt heating holder, which was manufactured by Gatan Inc, Model 652 Inconel. The heating temperature was monitored by a thermo-couple attached directly to the furnace. The heating trajectory of the experiment was plotted in red in Fig. S6. Gas injection pressure trajectory was plotted in blue in Fig. S6. Gas pressure was maintained at 15 mtorr except for a short period at 300 °C, where the pressure was raised to ~6.5 torr. The gun valve was closed during this process.

**Co valence determination.** The fitting method used in Figure 4 is only accurate when electro-megnetic optical condictions are the same when the references spectra and the intemediate spectra are recorded. That generally requires the reference spectra and the intemediate spectra are recorded in the same session because the dispersion and linearity of the spectrometer varies from time to time. In Figure 4, the results are self-referenced where the spectra from the two ends of the reaction are used as reference spectra as we know the materials can only be reduced. This method gives an accurate measurement of the reduction fraction. However, in Figure 5d, we do not have oxidized references taken at the same condition and therefore the L3/L2 method was used.

*Conflict of Interest*: The authors declare no competing financial interest.

*Acknowledgement*. This work was supported by the Office of Basic Energy Sciences, Chemical Science Division of the U.S. DOE under Contrast No. DE-AC02-05CH11231. We performed *ex situ* TEM experiments at National Center for Electron Microscopy (NCEM) of the Lawrence Berkeley National Laboratory (LBNL), which is supported by the U.S. Department of Energy (DOE) under Contract No. DE-AC02-05CH11231. The *in situ* environmental TEM experiments were carried out at the Center for Functional Nanomaterials, Brookhaven National Laboratory, which is supported by the U.S. Department of Energy, Office of Basic Energy Sciences, under Contract No. DE-AC02-98CH10886. EAP thanks Trevor Ewers and Prof. Paul Alivisatos for providing guidance and the access to the synthesis laboratory. HL Xin thanks Peter Ercius for helping with the tomography setup and Robert Hovden for the development of the Cornell e$^-$ Tomo reconstruction software. H Zheng thanks the funding support from DOE Early Career Research Program.

*Supporting Information Available*: Supplementary figures. This material is available free of charge *via* the Internet at http://pubs.acs.org.

FIGURES AND FIGURE CAPTIONS

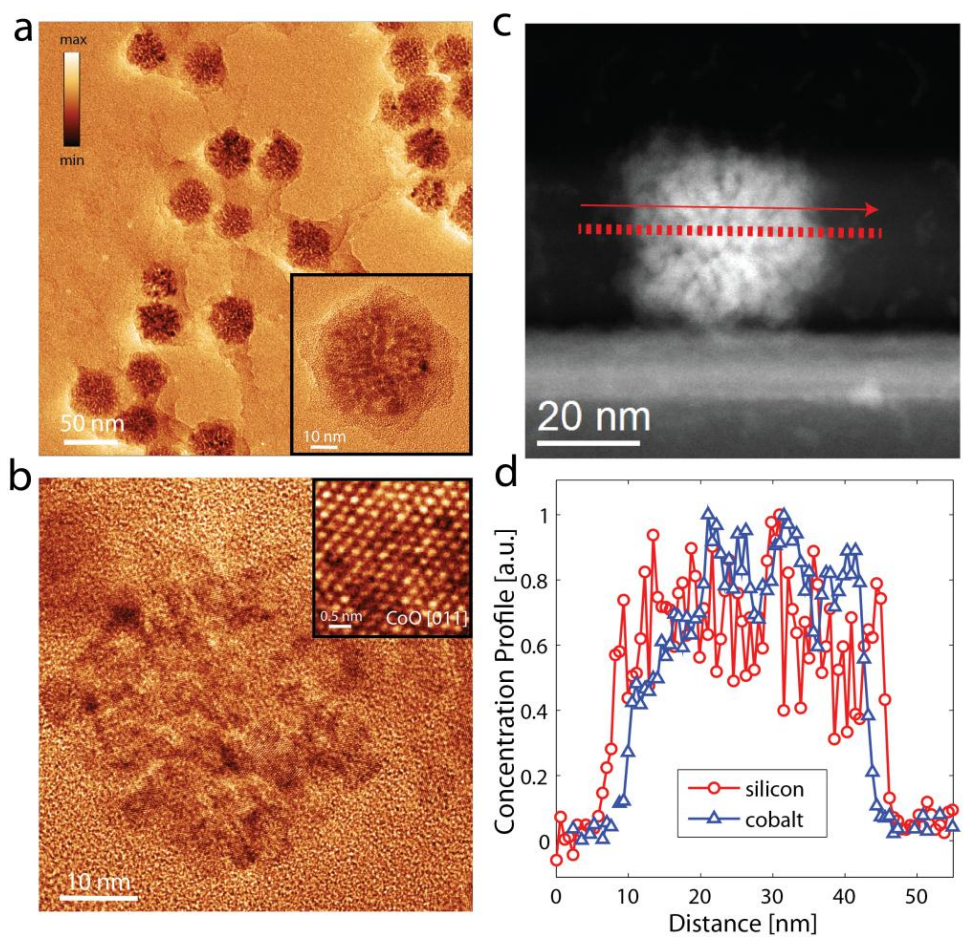

**Figure 1.** Morphology and composition of as-prepared $CoO_x/SiO_2$ nanocomposites. (a) Bright-field TEM image of the nanocomposites. (b) High-resolution TEM image of a $CoO_x/SiO_2$ Particle. (c) Annular dark-field scanning transmission electron microscopy (ADF-STEM) image of a $CoO_x/SiO2$ nanocomposite. The electron energy loss spectroscopic (EELS) line profile was recorded along the dashed line. (d) The extracted silicon and cobalt concentrations along the line profile in (e). The maximum intensity was normalized to one. Silicon concentration was integrated from the Si L2,3 edges. Cobalt concentration was integrated from the Co L2,3 edges.

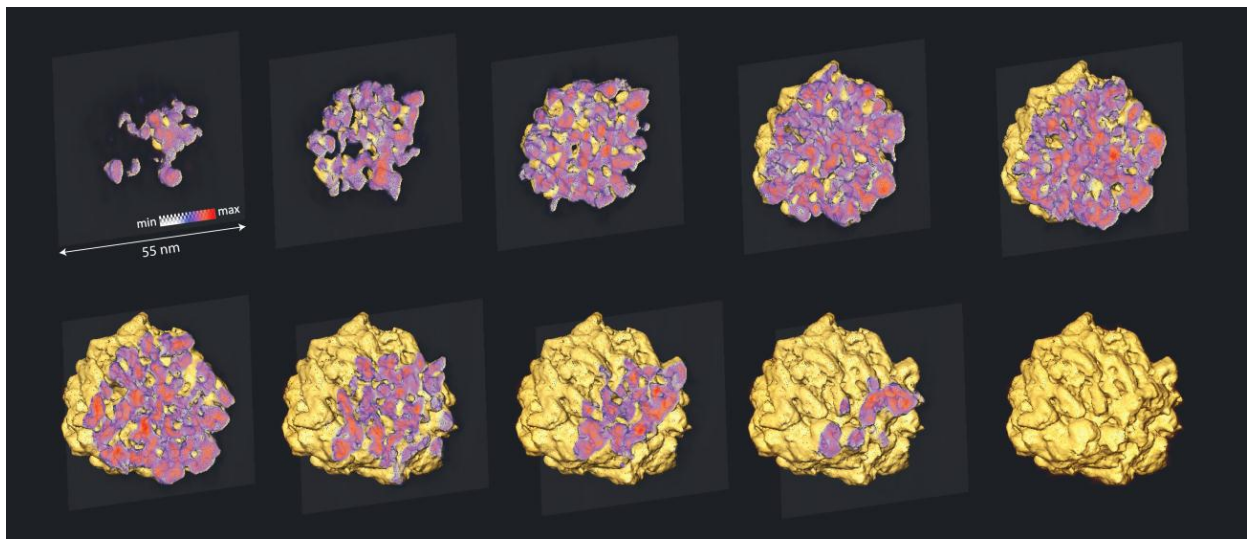

**Figure 2.** 3-D tomographic reconstruction of an as-prepared $CoO_x/SiO2$ nanocomposite. The progressing cross sections and the isosurfaces visualize the internal structures of the porous $CoO_x$ core.

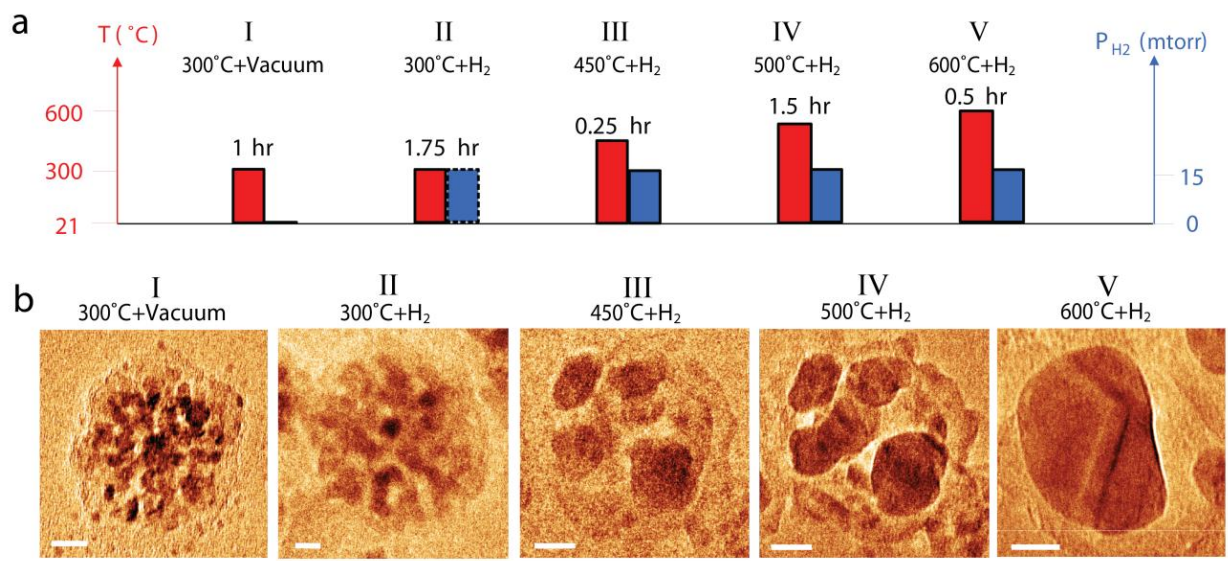

**Figure 3.** *In situ* observation of structural changes of Co/CoO$_x$ catalysts at different reduction conditions. (a) Heating and H$_2$ gas environmental conditions and their durations. (For a period of 30 minutes, the H2 gas pressure was raised to ~6.5 Torr. A detailed trajectory is shown in Figure S6.) (b) *In situ* TEM images of the nanocomposites under the corresponding environmental treatments. Images of large field of view are shown in supplementary Figure S8. (Scale bars are 10 nm.)

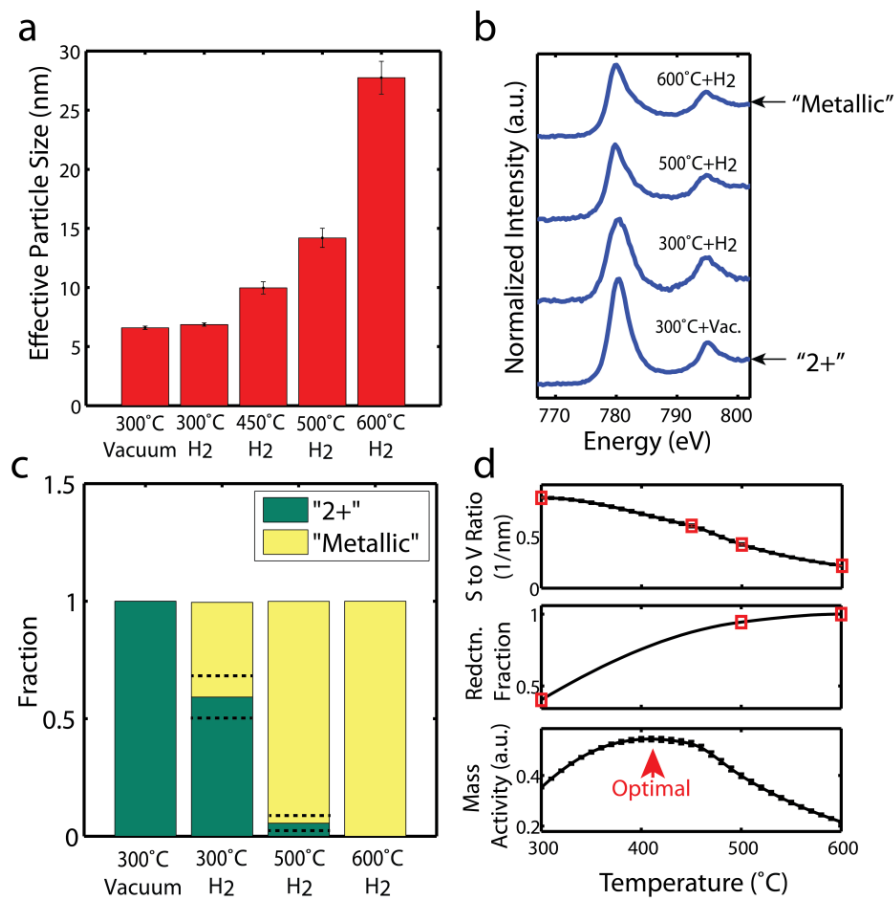

**Figure 4.** Correlation of valence state with coarsening (*via* effective particle size) in nanoporous Co/silica catalysts during $H_2$ reduction. (a) The effective particle size as a function of the reduction conditions. (b) Co $L_{2,3}$ edges of the catalysts, determined from *in situ* measurements. (c) The reduction fraction as a function of reduction conditions. The dashed lines mark 68% confidence intervals of the measurements. (d) Surface to volume ratio, fraction of reduced metallic cobalt calculated from (c) and (b), and the projected optimum condition.

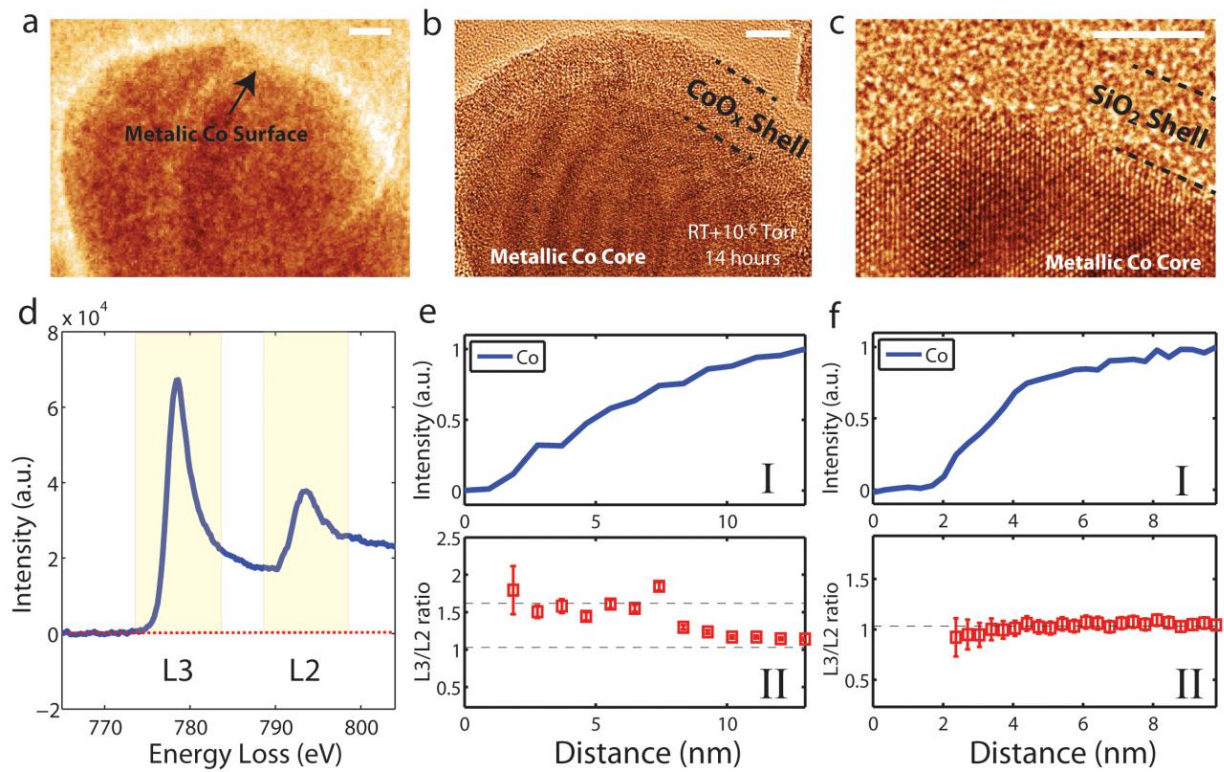

**Figure 5.** Comparison of silica coated and uncoated Co surfaces. (a and b) A Co particle with the exposed surface as indicated by the arrow. (a) The particle is fully metallic at 600 °C+$H_2$. (b) The exposed surface is oxidized 14 hours after reduction at room temperature in $10^{-6}$ Torr vaccum. (c) *Ex situ* TEM image of a reduced particle with an only 3.5 nm thick silica shell after being exposed to air at room temperature for more than three weeks. (d) The EELS spectral setup for L3/L2 ratio calculation. (e-f) *Ex situ* measurements of the L3/L2 ratio of across (e) the exposed surface and (f) the silica coated surface (see supplementary Figure S12 for the ADF-STEM images and the line profile positions). I: Co concentration normalized by the maximum intensity. II: the measured L3/L2 ratio as defined in (d). L3/L2 ratios are plotted for spectra with a normalized Co concentration higher than 0.1. (Scale bars are 4 nm.)

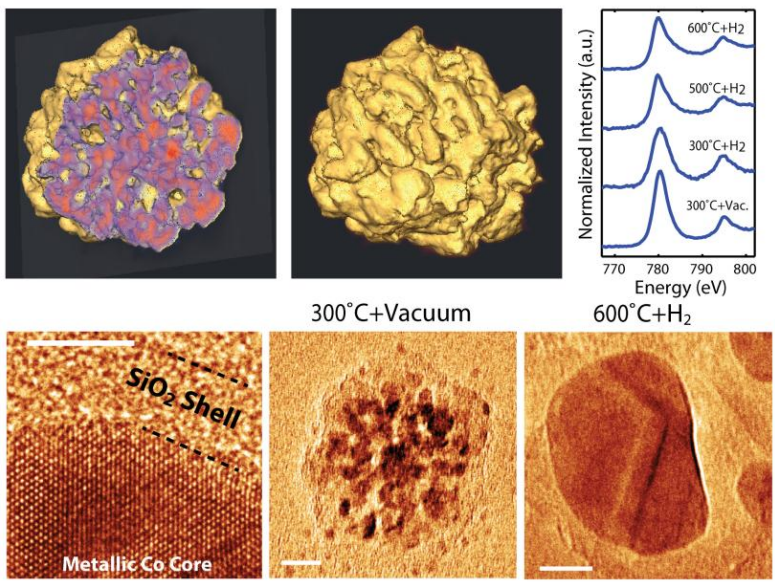

TOC Figure

# Supporting Information

## for

## Revealing Correlation of Valence State with Nanoporous Structure in Cobalt Catalyst Nanoparticles by *in situ* Environmental TEM

*Huolin L. Xin[1], Elzbieta A. Pach[1], Rosa E. Diaz[2], Eric A. Stach[2], Miquel Salmeron[1], Haimei Zheng[1]**

[1]Materials Sciences Division, Lawrence Berkeley National Laboratory, Berkeley, CA 94720, United States

[2]Center for Functional Nanomaterials, Brookhaven National Laboratory, Upton, NY 11973, United States

*Corresponding author: hmzheng@lbl.gov

This file contains: supplementary Fig. S1-S14.

# Supplementary figures

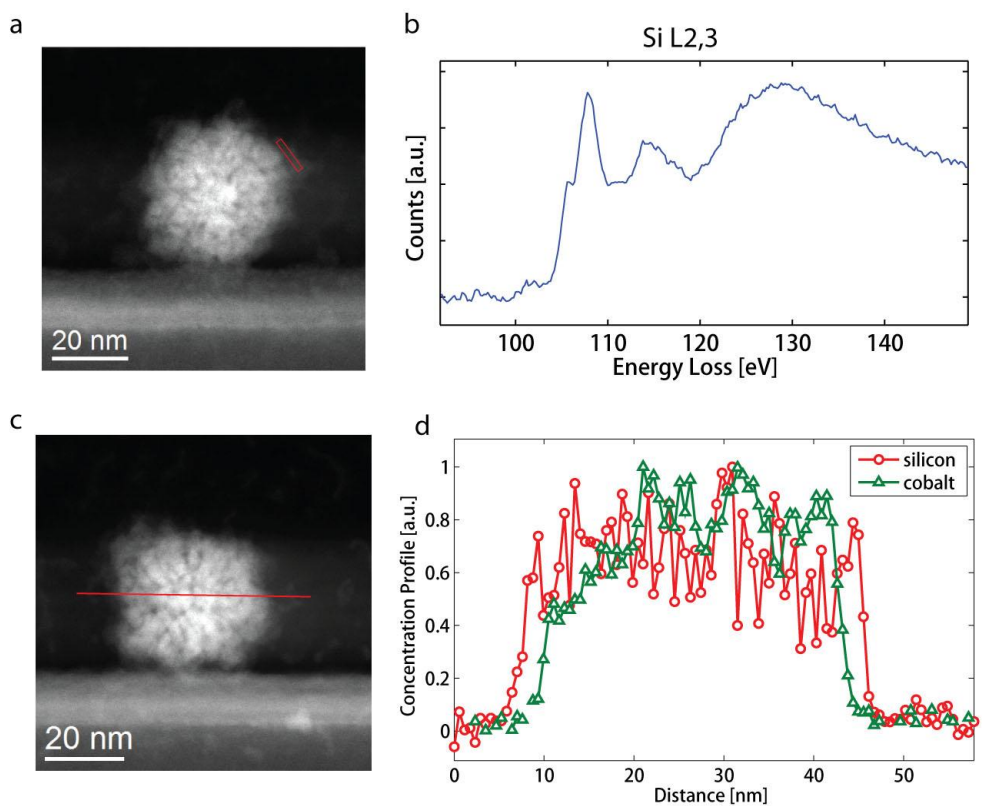

**Figure S1.** Spatially resolved STEM-EELS of the nanoporous Co/silica particles. (a-b) Localized probing of the electronic structure of Si in the silica. (a) An ADF-STEM image. The red box shows the location where the spectrum was acquired. (b) The near-edge fine structures demonstrate that the probed material is silicon dioxide. The spectrum shows the silicon L2,3 edges. (c-d) STEM-EELS line profile of Si and Co in the nanocomposite. (c) The red line shows where the atomic-size electron beam was scanned along. A hundred spectra were acquired successively from the left to the right along the line. (d) The normalized silicon and cobalt concentration across the particle. The fact that the spatial extent of cobalt is narrower than that of silicon demonstrates the silica is coated on the cobalt particle.

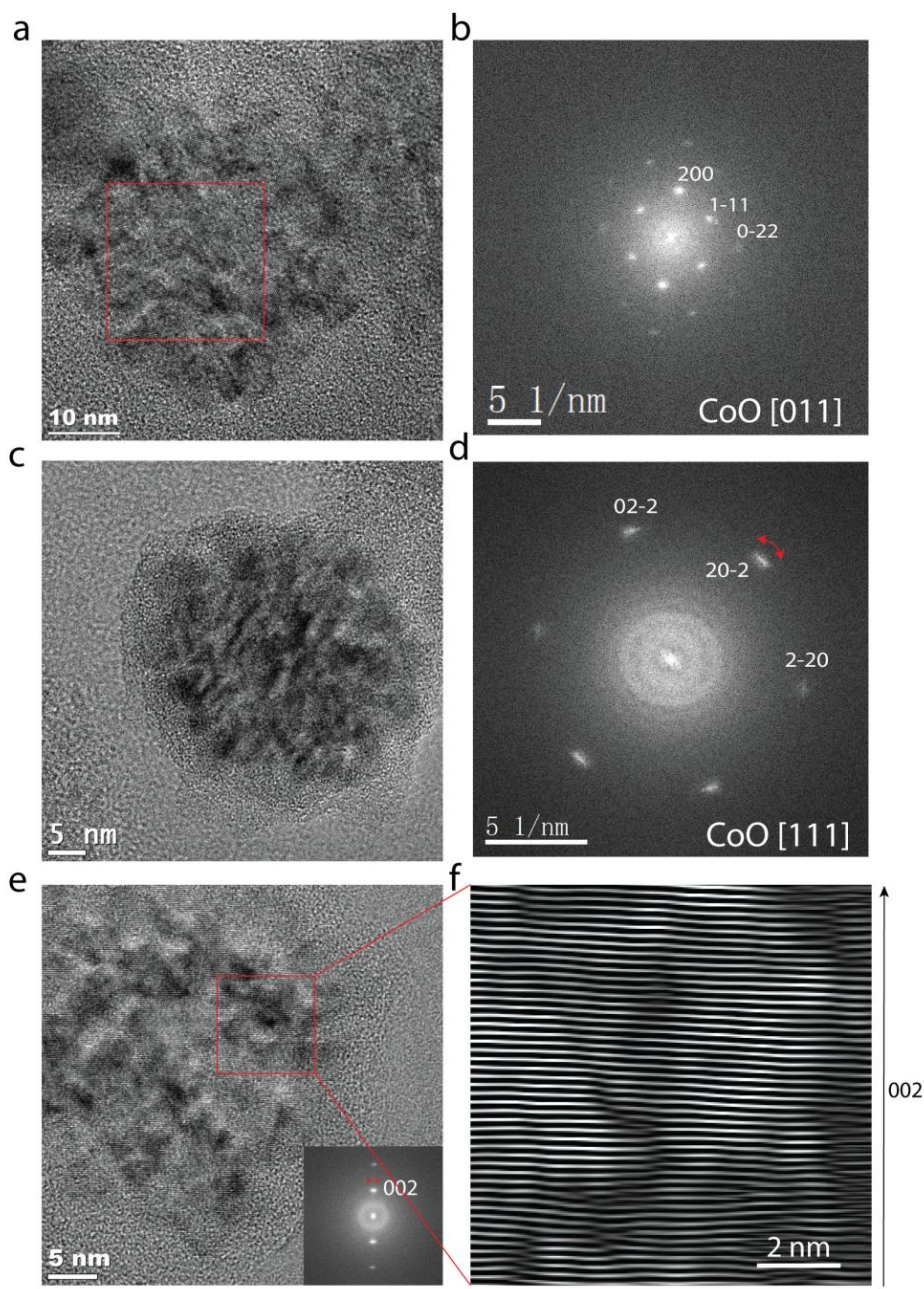

**Figure S2.** HRTEM imaging of the as-prepared CoO$_x$/SiO$_2$ nanocomposite. (a, c) HRTEM images and the corresponding (c, d) fast Fourier transform. (b, d) There is only one general set of orientations within a simple core. The angular spread of the reflection spots as indicated by the red arrow demonstrates that within the core, grains are relatively aligned with small angle grain boundaries. Zone axes are identified as the (b) CoO [011] and (d) CoO [111]. (e-f) HRTEM image of the core and its corresponding Fourier filtering analysis. (e) HRTEM image of an off-axis particle with strong 200 and 400 reflections. These two pairs of reflections are Fourier filtered to give the filtered image in (f). The filtered image highlights how (002) planes are reoriented at small-angle grain boundaries.

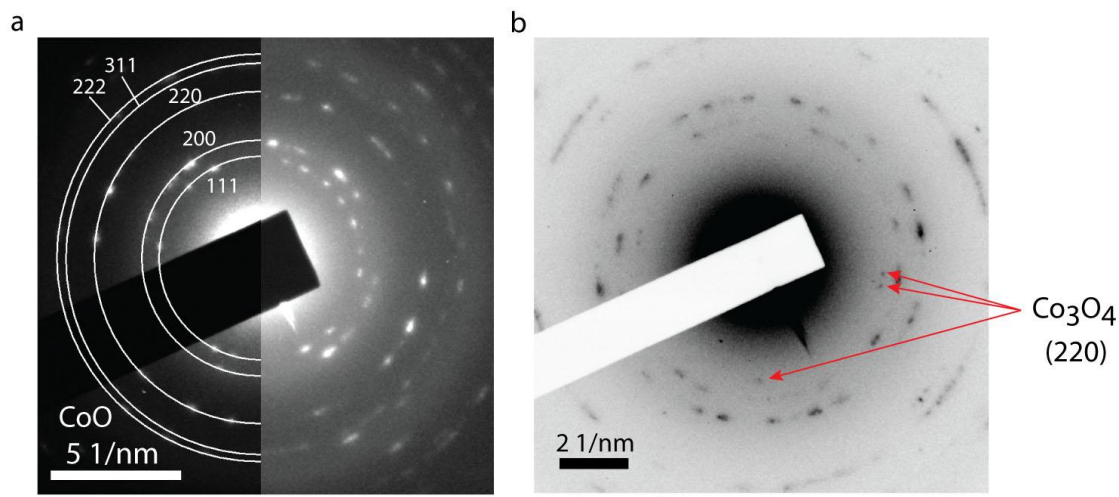

**Figure S3.** Selected area electron diffraction (SAD) pattern of the as-prepared $CoO_x/SiO_2$ nanocomposite. (a-b) SAD patterns of two different areas. (a) This area shows a pure CoO pattern. (b) This area shows a trace amount of $Co_3O_4$ as indicated by $Co_3O_4$ 200 reflections pointed by the red arrows.

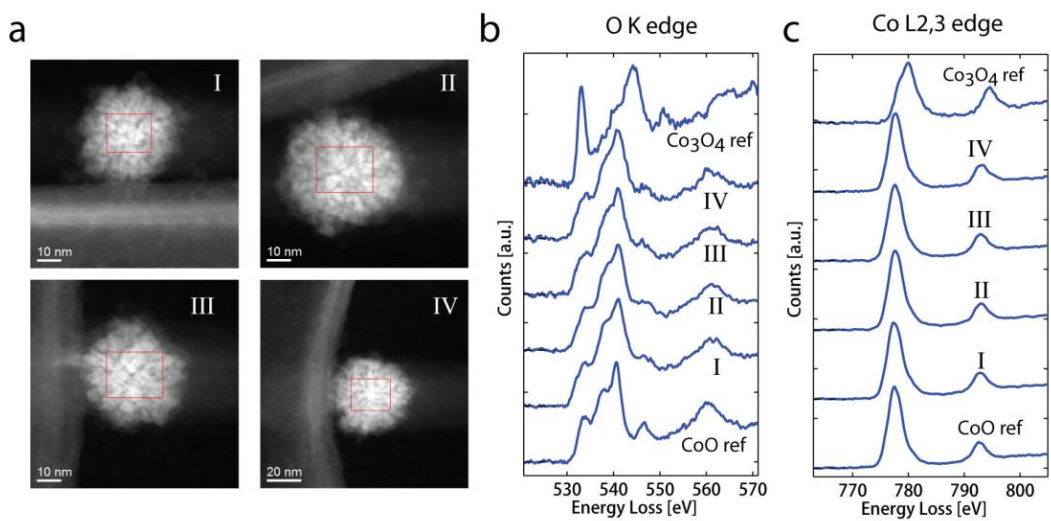

**Figure S4.** ADF-STEM images and EELS spectra of the as-prepared $CoO_x/SiO_2$ nanocomposites. (a) ADF-STEM Z-contrast images of four nanocomposites (I, II, III, IV). (b) The O K edge of the four nancomposites in comparison with the CoO and $Co_3O_4$ references. I, II, III, IV are almost identical to the CoO reference, which indicates CoO is the majority phase within the core. (c) The Co L2,3 edges of the nanocomposites in comparison with the CoO and $Co_3O_4$ references. The L3/L2 ratio of I, II, III, IV are almost identical to the CoO reference, which demonstrates the average valence of the cobalt core is very close to 2+.

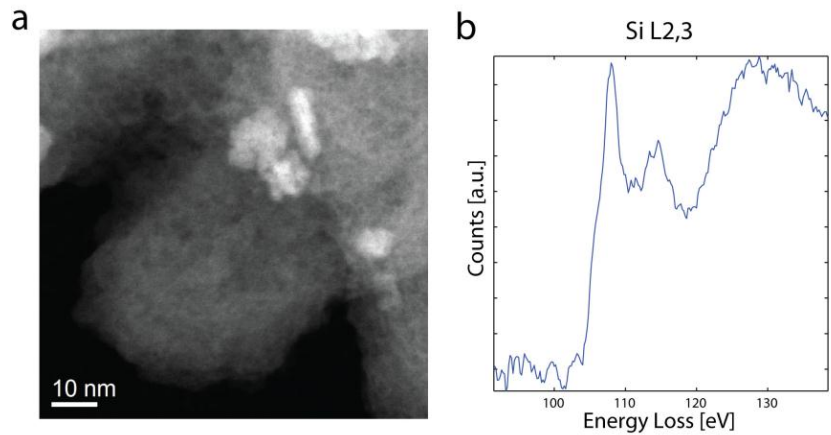

**Figure S5.** ADF-STEM imaging of the porous silica. (a) The ADF-STEM Z-contrast image of a different batch of sample with high silica/cobalt ratio. The speckles in the Z-contrast image shows there is variation in materials density along the beam incident direction, which is likely due to the presence of low-density pores. (b) The silicon L2,3 edges of the sample indicates the silicon in this porous silica is close to fully oxidized.

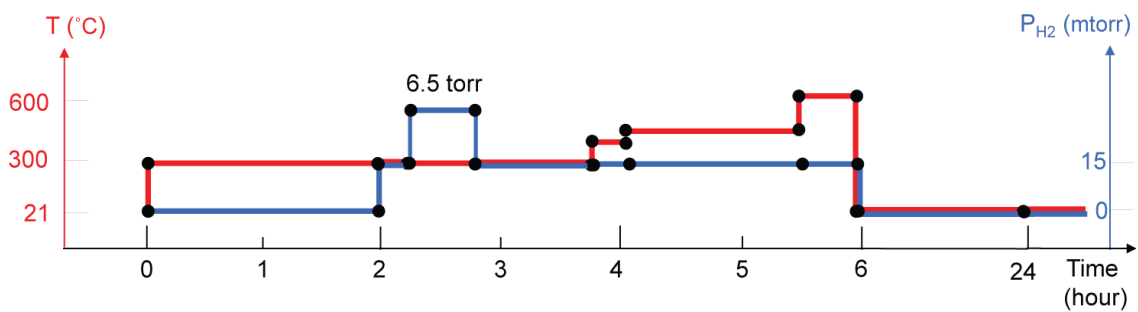

**Figure S6.** The detailed heating and gas trajectories as a function of time used in the *in situ* ETEM experiment.

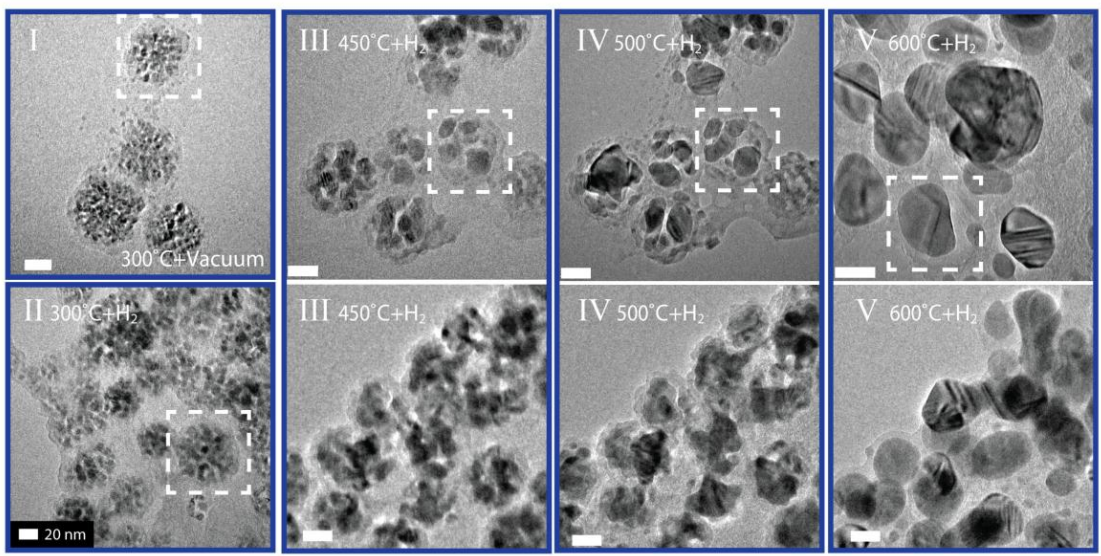

**Figure S7**. *In situ* observation of coarsening of Co/CoO$_x$ catalysts at different reduction conditions. Selected particles shown in Figure 3 are framed with dashed boxes.

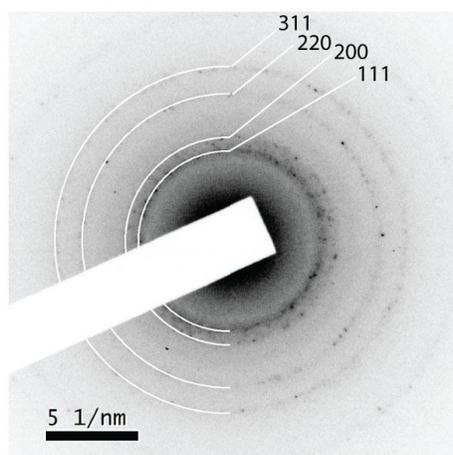

**Figure S8.** Diffraction pattern of the reduced sample. The various rings belong to metallic cobalt.

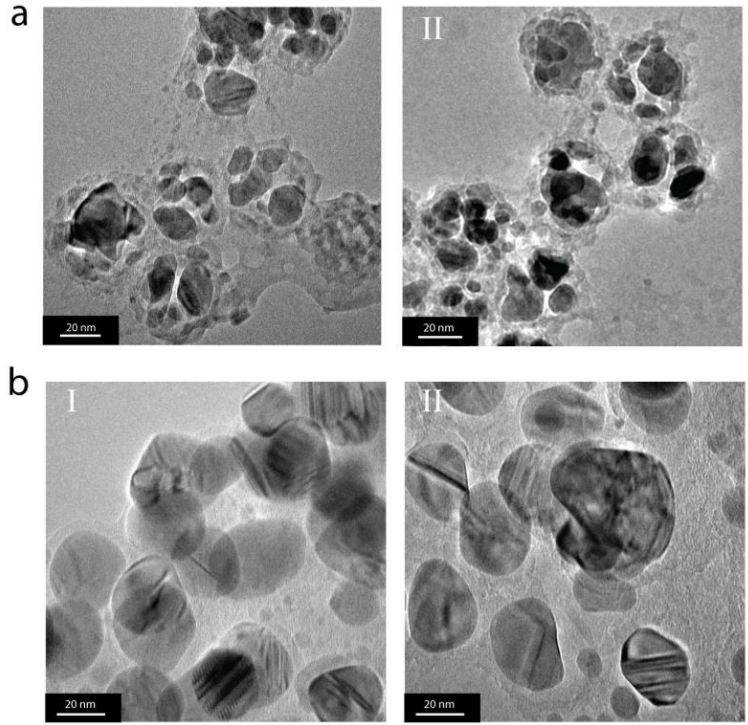

**Figure S9.** Comparison of the morphology of the CoO$_x$/Co cores in an e-beam illuminated area and a non-illuminated area. (a) 500 ℃+H$_2$. (I) Illuminated area. (II) Non-illuminated area. (b) 600 ℃+H$_2$. (I) Illuminated area. (II) Non-illuminated area.

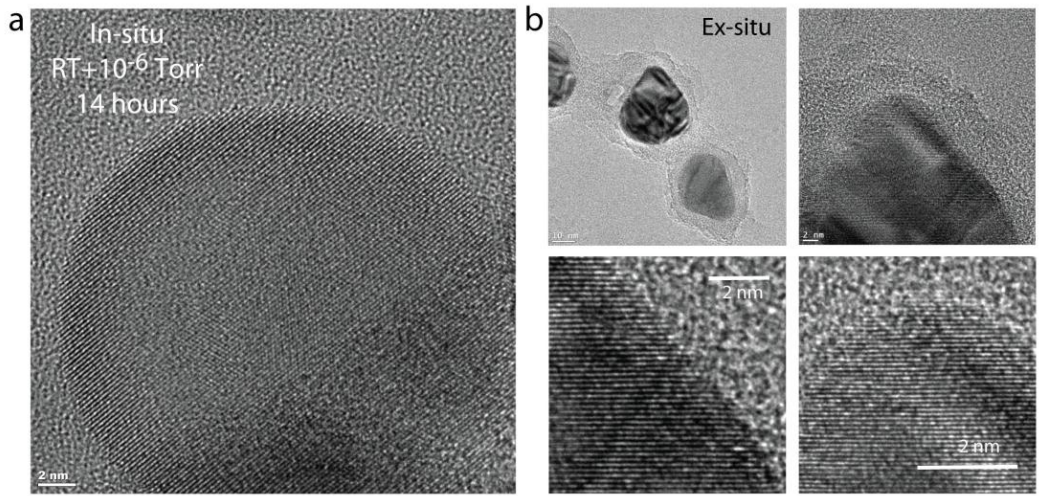

**Figure S10.** *In situ* and *ex situ* HRTEM images of the silica/cobalt interfaces. (a) An *in situ* image of a silica protected particle, which was taken 14 hours after the reduction. That the lattice was not disrupted at the silica/cobalt interface indicates the surfaces of the cobalt core are not oxidized. (b) An *ex situ* image of the ETEM reduced nanocomposite.

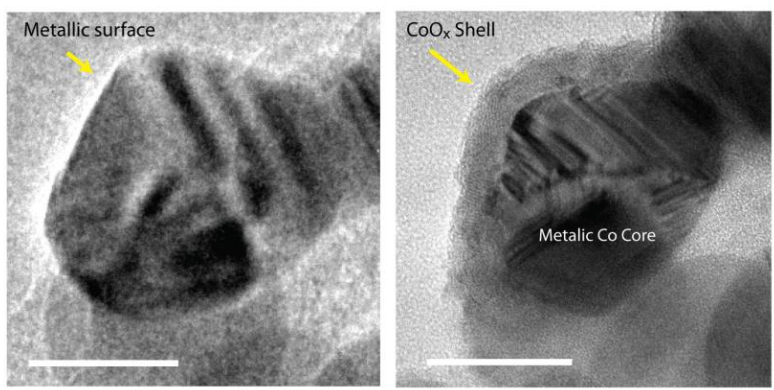

**Figure S11**. Uncoated Co surfaces are oxidized after exposed to air. (a) The reduced metallic particle in at 600 ℃+$H_2$. (b) *Ex situ* TEM observation of the same particle a week after the *in situ* experiment. Scale bars: 20 nm.

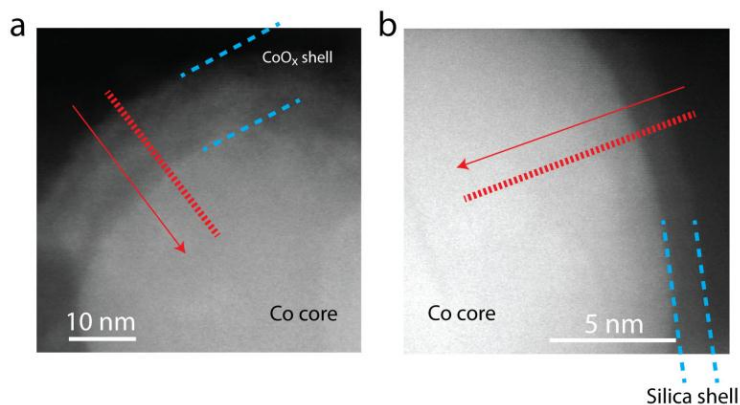

**Figure S12**. TEM and ADF-STEM image of cobalt particles with and without silica coating. The red dashed lines mark the positions where the EELS line profiles were recorded from. (a and b) The particle's top surface was oxidized due to the lack of silica coating (Fig. 5a). (c and d) The particle's surface is coated with a 2nm-thick silica layer.

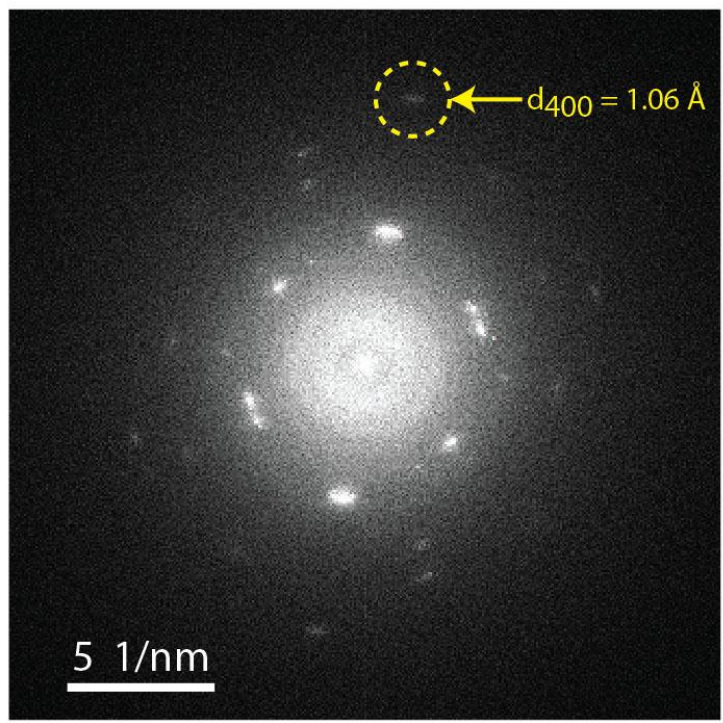

**Figure S13**. 2-D Fourier transform of Figure 1b. The information transfer is beyond 1.06 Ångstrom.

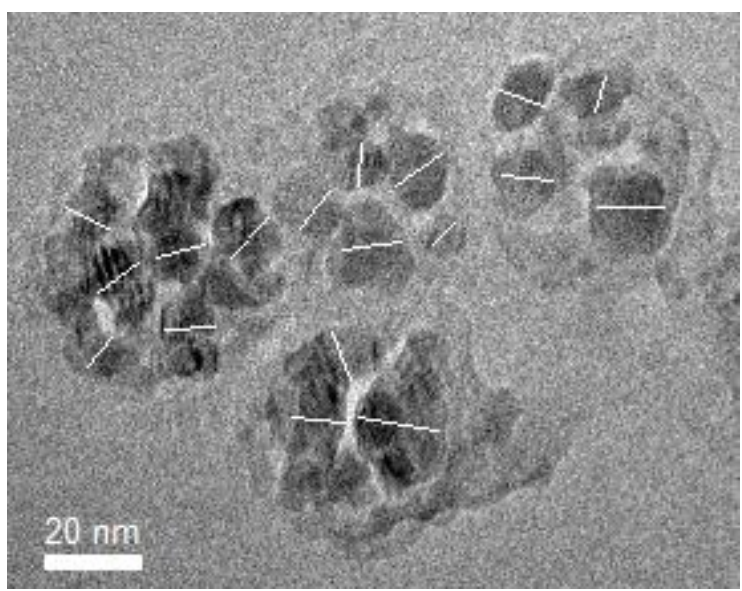

**Figure S14**. Extraction of the effective particle size from the *in situ* TEM images of the catalysts. The particle sizes are measured by manually placing lines over the major features in the catalyst particles. More than five individual Co/Silica nanocomposites are analyzed for each reaction condition.